# Nonlinear Dynamics of the Phonon Stimulated Emission in Microwave Solid-State Resonator of the Nonautonomous Phaser Generator[*]

D.N. Makovetskii

*A. Usikov Institute of Radio Physics and Electronics, National Academy of Sciences of Ukraine 12, Academician Proskura St., Kharkov 61085, Ukraine*

The microwave phonon stimulated emission (SE) has been investigated in a solid-state resonator of a non-autonomous acoustic quantum generator (phaser). Branching and long-time refractority (absence of the reaction on the external pulses) for deterministic chaotic and regular processes of SE were observed in the experiments with weak and strong electromagnetic pumping. The clearly depined increase of the number of the independently co-existing SE states as the pumping level rise has been observed both in physical and computer experiments. This confirms the analytical estimations of the branching density in the phase space. The nature of the SE pulses refractority is closely connected with such branching and reflects the crises of strange attractors, i.e. their collisions with unstable periodic components of the higher branches.

The stimulated emission (SE) of microwave phonons has been first observed in experiments on quantum hypersound amplification, when in 1960-ies it was reported for dielectric single crystals doped with paramagnetic ions of the iron group [1] - [3]. The hypersound amplification is due to inversed population of the ion spin levels, transitions between which are allowed for the spin-phonon interaction, and is, in fact, an acoustic analog of the linear maser amplification of electromagnetic fields [4], [5].

At the same time, the nature of this effect – quantum *generation* of microwave phonons – has remained not fully understood for a long time. There was some inertia of thinking, with analogies proposed [6] between the photon quantum generator (phaser) and electromagnetic maser generator in the same way as it had been done for amplifiers.

Experimental studies of phonon SE in $Ni^{2+}:Al_2O_3$ and $Cr^{3+}:Al_2O_3$ crystals ([7] - [10]) have shown that phaser generation mechanism is actually much closer to the processes in optical lasers than those in maser generators. Really, due to very low velocity of hypersound (as compared with the velocity of light) the wavelength of acoustic SE in a microwave solid-state resonator is about 1-3 μm, i.e., it corresponds to the near-infrared range of electromagnetic radiation. The resonator Q-factor $Q_{CAV}$ is, like in typical lasers, $Q_{CAV} \approx 10^6$, i.e., again it is by several orders of magnitude higher than in electromagnetic masers. As a result, observed SE power spectra, transient processes, stability of stationary modes and other properties of a phaser are very much similar to these characteristics for solid-state lasers at $T_{1S} \gg T_{CAV} \gg T_{2S}$. Here $T_{1S}, T_{2S}$ are, respectively, axial and transversal relaxation times of active (inverted) centers, $T_{CAV} = Q_{CAV}/\Omega_{CAV}$, where $\Omega_{CAV}$ is the operation mode frequency of the resonator with active medium, and $\Omega_{CAV} \approx \Omega_S \approx \Omega_{SE}$, $\Omega_S$ is the quantum transition frequency for inverse population difference, $\Omega_{SE}$ - SE carrier frequency.

However, there is also a fundamental difference between phaser and optical lasers, which is related to the intensity of intrinsic quantum noises (intrinsic, or spontaneous, emission) $J_{spont}$. As for a phaser $\Omega_S = 3 - 10$ GHz, i.e., five orders of magnitude less than for a laser, the relative level of the intrinsic component in the first generator is by ~15 orders lower than in the second one (due to $J_{spont} \propto \Omega_S^3$). In fact, this allows us to consider a phaser as a *deterministic* dynamic system practically in all practically available ranges of SE intensity.

Investigation of complex (including irregular) determined movements in dynamic systems is now steadily proceeding. This is related not only to the fundamental character of this problem, but also to prospects of practical application of the methods involved and results obtained in studies on the mechanisms of associative memory [11], [12], for development of computer image and symbol recognition [13], [14] etc. The multiplicative noises (which also include intrinsic radiation) affect the behavior of dynamical systems in a very non-trivial manner [15], e.g., leading to roughing of phase space topology [16] and substantial reduction of the associative memory volume. Such considerations substantially increase our interest in phaser generator-type systems, where, as distinct from optical lasers, internal noise level is low for practically all feasible combinations of controlling parameters (CP).

In the present paper, we summarize the data of our experimental studies of a non-autonomic phaser generator, which were partially reported in our separate presentations [17, 18]. We also report numerical modeling of SE dynamics on the basis of a deterministic generation model that is an acoustic analog of the laser model [19]. The experiments were carried out using a ruby phaser as described earlier [7] - [10] under modulation of various CP (pumping, static magnetic field $H$, intensity of hypersound $J_{inj}$ injected into the resonator, etc.) in the frequency range $\omega_m = 30 - 3000$ Hz.

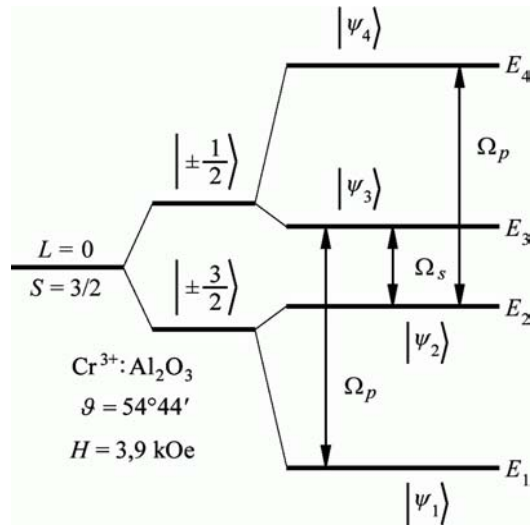

Fig. 1. Energy level diagram for $Cr^{3+}$:$Al_2O_3$ active crystal.

The $Cr^{3+}$:$Al_2O_3$ spin system is formed upon splitting of the principal energy level of $Cr^{3+}$ ion with orbital quantum number $L = 0$ and spin $S = 3/2$ in a trigonal crystalline field and static magnetic field $H$ (Fig.1). Electromagnetic pumping of the $Cr^{3+}$:$Al_2O_3$ spin system was carried out with a klystron of power $P_{pump} = 12$ mW at frequency $\Omega_P = \Omega_{P0} + \Delta\Omega_P$ using a cylindrical cavity on the $H_{011}$ mode. The cavity was tuned to frequency $\Omega_{P0} = 23$ GHz. This corresponds to the resonance magnetic field $H_0 = 3,9$ kOe at the angle $\vartheta = 54°44'$ between $\vec{H}$ and the third order crystallographic axis $\vec{C}$ of ruby (when conditions for symmetrical pumping are realized – see Fig.1). Detuning of the pumping source $\Delta\Omega_P$ varied within several tens of MHz. A solid-state microwave Fabry-Perot acoustic resonator (FPAR) of Q-factor $Q_{CAV} \approx 5 \cdot 10^5$ ($T_{CAV} \approx 10^{-5}$ s) was placed inside the pumping cavity along its axis. FPAR is a cylinder of synthetic rose ruby with optically planar and parallel end sides – acoustic mirrors.

Upon one of the FPAR acoustic mirrors, deposition a thin textured ZnO film with Al sublayer was applied by vacuum evaporation, this film being the main element of the hypersound converter. Using this film, detection was carried out of SE signals having carrier frequency $\Omega_{SE}$ = 9,16 GHz and intensity $J(t)$. The SE signal is formed on $E_2 \leftrightarrow E_3$ transition of Cr3+ ions under electromagnetic pumping of $E_1 \leftrightarrow E_3$ и $E_2 \leftrightarrow E_4$ transitions (diagram of energy levels for $Cr^{3+}$:$Al_2O_3$ is presented in Fig.1, spin levels $E_i$ and respective wave functions $|\psi_i\rangle$ are numbered in the order of increasing energy). By means of the same film, external longitudinal hypersound of frequency $\Omega_{inj} \approx \Omega_S \approx$ 9 GHz and pulse intensity $J_{inj} \leq 300$ mW/cm$^2$ was injected into the ruby crystal. All experiments were carried out at 1.8 K. The axial relaxation time on the active transition was $T_{1S} \approx 0,1$ s, the transverse relaxation time $T_{2S} \approx 10$ ns.

The third-order axis $\vec{C}$ of ruby was coincided with FPAR geometrical axis $\vec{O}$, and conical refraction was observed for the transverse component of the phonon SE[7]; thus, the contribution of the longitudinal component to SE intensity $J$ was overwhelming. As the texture axis of ZnO film was also parallel to axes $\vec{C}$ and $\vec{O}$, the injected hypersound was also purely longitudinal. It should be stressed that the momentary value $J(t)$ is determined by SE processes in FPAR and is (alongside with the inverse population difference $n$) a dynamic variable, while $J_{inj}$ is among CP. The time-averaged SE intensity value $\langle J \rangle$ can be different from zero only if $P_{pump} > P_{pump}^{(G)}$. Here $P_{pump}^{(G)}$ is the value of electromagnetic pumping power at frequency $\Omega_P = 23$ GHz, where phaser generation begins. The pumping parameter $A$ (the ratio of inverted population difference at frequency $\Omega_S \approx 9$ GHz to its threshold value) is equal to unity.

At low levels of pumping ($A - 1 \approx 1/30$) and small periodic perturbation of the system, ($k_m \ll 1$, where $k_m$ is modulation coefficient of one of CP, e.g., $P_{pump}$, $H$ or $J_{inj}$) regular modulation of the recorded SE signal $J$ with period $T_m$ is always observed in experiments all over the modulation frequency range $\omega_m = 30 - 3000$ Hz. This corresponds to a soft birth (emergence) of a limiting cycle of period $T_m$ on the main (zeroth) generation branch $B_0$. When $k_m$ is increased, the SE output signal modulation coefficient smoothly rises up to values normally not higher than 60 - 70 %. Then in a certain critical point $k_m = k_m^{(0)}$ the value of $R_{SE}$ increases stepwise to nearly 100 %, not acquiring, however, a pulse character and preserving the same SE period $T_m$. If we change the sign of $dk_m/dt$, hysteresis will be observed of the $R_{SE}(k_m)$ dependence – the jump of $R_{SE}$ back (decreasing) is at $k_m^{(1)} < k_m^{(0)}$, indicating the existence of an above-located hard branch $B_1$ of IR.

Thus, even in the range of $k_m$ where SE is not yet of pulse character, a bistability of periodic movements is observed. Co-existence of the soft $B_0$ and hard $B_1$ branches with one and the same modulation period of integral intensity of the generated photon flux is realized. If now, being on the branch $B_1$, to increase (and not to decrease) $k_m$, at $A - 1 \approx 1/30$ consequent doubling of the SE period will be observed according to Feigenbaum's scenario, ending in the transition to the chaotic pulse modulation mode $J(t)$. Varying the system detuning over magnetic field $\Delta H = H - H_0$ within several

Oersteds, we succeeded in finding two qualitatively different types of irregular (chaotic) modulation with pseudoperiods $T_m$ and $2T_m$ (pseudoperiod is the maximum interval between the neighboring pulses rounded to the nearest integer value [20]).

As for the soft branch $B_0$, at $A-1 \approx 1/30$ it has strictly regular character with period $T_m$ in the all range of $k_m \in \left(0, k_m^{(0)}\right)$, cutting itself off at the transition point to the hard branch $B_1$ independently of the $\Delta H$ detuning (though specific values of $k_m^{(0)}$ naturally change upon variation of $\Delta H$).

As it was shown by experiments with modulation of each of the selected CP ($P_{pump}$, $H$, $J_{inj}$) the further increase in $k_m$ at low pumping level ($A-1 \approx 1/30$) leads to appearance of new hard branches $B_2$, $B_3$, ..., which begin by an appropriate periodic (which is now pulse) SE mode with period $MT_m$, where, respectively, $M = 2, 3, ...$ , and most often end by the chaotic mode after the series of Feigenbaum's doublings. It should be noted that coexistence region of the SE branches was typically narrow for small $A-1$, and at $\Delta H < 4$ Oe no bistability of periodical movements with $M > 1$ was observed at all.

The picture becomes qualitatively different when pumping parameter $A$ is increased up to values of $A \approx 1.5 - 1.8$. Here strong overlapping of periodical regions occurs on branches $B_2$, $B_3$, ..., accompanied by a hysteresis under slow scanning of any of non-modulated CP (including $k_m$, $\Delta\Omega_p$, etc.). Fig.2 shows a typical picture observed for coexistence of the regular SE modes.

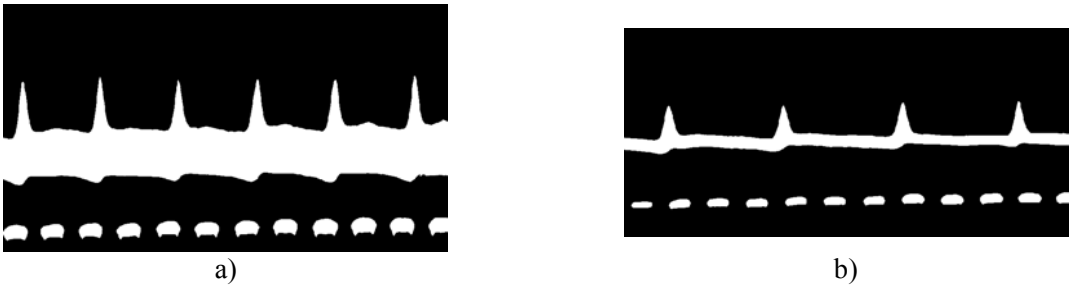

a) b)

Fig. 2. Regular SE of microwave phonons (experiment): coexistence of the modes
with doubled (a) and tripled (b) sequence periods of generation pulses.

The SE modes with doubled (Fig.2a) and tripled (Fig.3b) sequence periods of generated phonon pulses are coexisted for common modulation period $T_m$ and a fixed set of other CP. Transitions $2T_m \leftrightarrow 3T_m$ are of hysteretic character when $\Delta\Omega_p$ is scanned within about 1 MHz. The dashed line indicated the pumping modulation period $T_m = 1/120$ s. The amplification factor of the recording equipment in Fig.2a is four times larger than in Fig.2b.

It should be underlined that the observed movement with period $2T_m$ (Fig.2a) is not an intermediate stage of the Feigenbaum's scenario on the first hard branch, but is a primary SE mode for the second hard branch – just as the primary mode with period $3T_m$ for the third hard branch (Fig.2b). This follows from the fact that at $dk_m/dt < 0$ both for $B_2$ and $B_3$ SE frustration (suppression) points $k_{m\downarrow}^{(2)}$ and $k_{m\downarrow}^{(3)}$ were found experimentally, where after sign inversion of $dk_m/dt$ (but with other CP *unchanged*) the disrupted movements were not restored. It was possible to restore the modes $2T_m$ (Fig.2a) and $3T_m$ (Fig.2b) only by stepwise variation of other CP – the most convenient was to apply a hypersound monopulse $J_{inj}$ up to 10 mW/cm$^2$ at $k_{m\uparrow}^{(2,3)} = k_{m\downarrow}^{(2,3)} + \Delta k_m^{(2,3)}$. It appeared that restoration occurred even at $\Delta k_m^{(2,3)} \approx 10^{-2} k_{m\downarrow}^{(2,3)}$, with further increase in $J_{inj}$ not leading to new modes for such low overcriticities.

It follows from these experimental data that at $k_m \approx k_{m\uparrow}^{(2)}$ the second hard branch, which is shown in Fig.2a, is also the toppest, i.e., located above the others. Similarly, at $k_m \approx k_{m\uparrow}^{(3)}$ it is the third hard branch that is the toppest (Fig.2b). From the other side, the fact that $k_{m\downarrow}^{(2,3)} \approx k_{m\uparrow}^{(2,3)}$ directly indicates that the periods $2T_m$ and $3T_m$ for these branches are the initial ones. In other words, SE refractority (i.e., absence of SE pulse during time $T_R = KT_m$, where $K > 1$) for hard branches $B_2$ and $B_3$ appears already at the moment of their birth. The presence of refractority is the main qualitative difference between the mode with period $2T_m$ on branch $B_2$ from the mode with the same period $2T_m$ (after Feigenbaum's doubling) on branch $B_1$. Really, for the latter mode doubling of the period ($T_{SE} = 2T_m$) is accompanied only by changes in the amplitude ratio of the neighboring SE pulses that follow with a unit pseudoperiod $T_R = T_{SE}/2 = T_m$ (no refractority). At the same time, for branch $B_2$ we have $T_R = T_{SE} = 2T_m$.

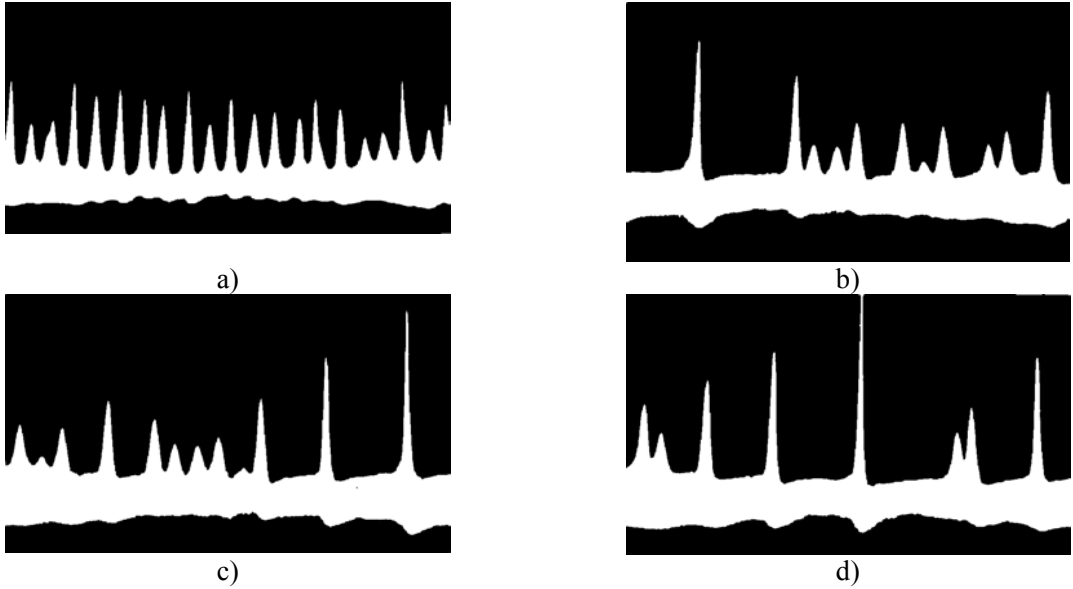

Fig. 3. Chaotic SE of microwave phonons (experiment): one realization with constant refractority (a) and three realizations with variable refractority (b)

Another type of refractority is observed at large $\Delta\Omega_P$, when, in particular, the effective pumping parameter is also decreasing. Fig.3 shows SE oscillograms recorded at $\Delta\Omega_H = 1$ Oe, $\omega_m = 120$ Hz in the cases of detuning $\Delta\Omega_p \approx 4$ MHz (oscillogram a) and $\Delta\Omega_p \approx 30$ MHz (oscillograms b-d). A transition takes place from the so-called helical chaotic mode with pseudoperiod $T_R = 1T_m$ (when each period of pumping modulation $T_m$ is matched by one SE pulse) to the chaotic mode with variable refractority, where pseudoperiods $T_R = KT_m$ ($1 \leq K \leq 5$) alternate irregularly. Amplification factor of the recording equipment in Fig.3 is the same for all oscillograms a-d. Besides large refractority intervals that appear on oscillograms b-d, sharp increase in the amplitude of pulses (as compared with oscillogram a) is observed, which is an evidence of substantial expansion of the attraction range of the corresponding attractor in the phase space of the system.

Subsequent measurements have shown that refractority of variable duration like shown in Fig.3 (oscillogrammes b-d) are observed for chaotic SE pulse modes even more often that purely helical chaotic movements with the unit half-period shown in Fig.3a. It is essential that the observed transitions between SE chaotic modes with different maximum half-periods be of non-hysteretic character, as distinct from the above-described hysteretic transitions between independent regular branches with different periods of movement (like those shown in Fig.2). To clear up specific mechanisms of branching of SE modes and appearance of variable refractority in chaotic realizations of pulse phonon generation, we should address the deterministic SE model in a high Q-factor resonator with active centers, for which the condition $T_{1S} \gg T_{CAV}$ is also met (this model is an acoustic analog of the deterministic model for class "B" lasers [19]). In this paper, we will limit ourselves in our numerical modeling by a relatively simple case of a three-level active system. Although we used experimentally a more complex four-level system (Fig.1), which allowed us to substantially enhance the inversion.

Let us consider a three-level spin system $E_3 > E_2 > E_1$, where signal transition $E_1 \leftrightarrow E_2$ (S-transition) is allowed for interaction with a coherent microwave acoustic field (hypersound) of specified direction and polarization, and the pumping transition $E_1 \leftrightarrow E_3$ is allowed for magneto-dipole interactions with the corresponding pumping field. The third transition (F-transition) is an idle (no-load) one, and its frequency $\Omega_F \equiv (E_3 - E_2)/\hbar$ is, by definition, not equal to the frequency of S-transition $\Omega_S \equiv (E_2 - E_1)/\hbar$ or its multiple values. In addition, $\Omega_F$ is not an integer divisor of the P-transition frequency $\Omega_P \equiv (E_3 - E_1)/\hbar$.

In the impurity paramagnetic $Cr^{3+}$:$Al_2O_3$, where the ground state (orbital quantum number $L = 0$, spin $S = 3/2$ – see Fig.1) is weakly bonded with the lattice (just by spin-orbital interaction), longitudinal relaxation times $T_{1S}, T_{1P}, T_{1F}$ for all the above-mentioned transitions at low temperatures are many orders of magnitude higher than the respective latitudinal relaxation times $T_{2S}, T_{2P}, T_{2F}$. Consequently, it is possible to choose the amplitude of microwave electromagmetic pumping field $H_{1P}$ such as to make populations $n_1$ и $n_3$ of spin levels $E_1$ equal, but with broadening of these levels still the same as at $H_{1P} = 0$. In other words, we assume that the following two inequalities hold simultaneously:

$$Z_P \gg 1; \quad (T_{2P}/T_{1P})Z_P \ll 1 \tag{1}$$

or, in a somewhat different form

$$Y_P T_{1P} \gg 1 \gg Y_P T_{2P}, \tag{2}$$

where $Z_P$ is saturation factor of P-transition

$$Z_P = \frac{1}{4} T_{1P} T_{2P} \gamma_P^2 H_{1P}^2, \tag{3}$$

and $Y_P$ is interaction probability of the pumping field with the P-transition. Here $\gamma_P$ is effective gyromagnetic ratio for this spin transition (accounting for direction and polarization of vector $\vec{H}_{1P}$).

Besides this, phonon life time $T_{CAV}$ in an FPAR of high Q-factor does normally meet the

requirement

$$T_{1S} \gg T_{CAV} \gg T_{2S}. \tag{4}$$

Similar inequalities are true for photon life times in a microwave electromagnetic pumping resonator. In addition, it is usually assumed that the pumping resonator has no intrinsic frequencies in the vicinity of $\Omega_F$.

In this case, to calculate the difference in populations $\Delta N \equiv n_2 - n_1$ on S-transition, one could use the balance approximation equations, which, accounting for equality $n_3 = n_1$, can be presented in the following form:

$$\begin{cases} \dfrac{dn_2}{dt} = W_{12}n_1 + W_{32}n_1 - W_2 n_2 - Y_S \Delta N \\ \dfrac{dn_1}{dt} = W_{21}n_2 + W_{31}n_1 - W_1 n_1 + Y_S \Delta N \end{cases}, \tag{5}$$

where $W_2 \equiv W_{21} + W_{23}$; $W_1 \equiv W_{12} + W_{13}$; $Y_S$ is the probability of interaction of the hypersound field with the spin S-transition; $W_{ij}$ are probabilities of longitudinal spin relaxation. Following [1], [7], we obtain

$$Y_S = T_{2S} q_u^2 U^2 |F_{12}|^2 / \hbar^2 \equiv Z_S / T_{1S}, \tag{6}$$

where $q_u = \Omega_S / V_u$; $U$ and $V_u$ are, respectively, the hypersound amplitude and phase velocity; $F_{12}$ is spin-phonon interaction factor for S-transition at specified values of the hypersound polarization and propagation direction. General expressions for $F_{12}$ can be found in [1]. For the case of latitudinal hypersound propagating along the crystallographic axis $\vec{O}$ of the third or higher order (as it is in the ruby phaser at 9 Ghz [7]), we have from [1]

$$F_{12} = \frac{\partial}{\partial \varepsilon_{zz}} \langle \psi_1 | \hat{H}_{su} | \psi_2 \rangle. \tag{7}$$

Here $\varepsilon_{zz}$ is a component of elastic deformation tensor; $\psi_1$ и $\psi_2$ are wave functions corresponding to spin levels $E_1$ and $E_2$; $\hat{H}_{su}$ is the spin-phonon interaction Hamiltonian.

Using the approximation $\hbar \Omega_P \ll 3 k_B \theta$ (where $k_B$ is the Boltzmann constant; $\theta$ - temperatute of the thermostat), we find from (5)

$$\frac{d(\Delta N)}{dt} = -2(\Delta N) Y_S + \frac{(\Delta N)_0 - (\Delta N)}{T_{1S}^{(P)}}, \tag{8}$$

where

$$(\Delta N)_0 = (-4 s_\theta W_S + 2 f_\theta W_F - 2 p_\theta W_P) W_E^{-1} N_c;$$

$$T_{1S}^{(P)} = 3 / W_E; \quad N_c = n_1 + n_2 + n_3;$$

$$s_\theta = \hbar \Omega_S / 3 k_B \theta; \quad W_S = \frac{W_{12} + W_{21}}{2};$$

$$p_\theta = \hbar\Omega_P / 3k_B\theta\,; \quad W_P = \frac{W_{13} + W_{31}}{2}\,;$$

$$f_\theta = \hbar\Omega_F / 3k_B\theta\,; \quad W_P = \frac{W_{23} + W_{32}}{2}\,;$$

$$W_E = (6 + 2s_\theta)W_S + (3 - f_\theta)W_F + 2p_\theta W_P.$$

The effective longitudinal relaxation time (relaxation time of active centers for the phaser signal channel) $T_{1S}^{(P)}$ that is referred to in (8) is not the conventional spin-lattice relaxation time $T_{1S}^{(0)}$ used in studies of passive systems, because pumping ("hidden" in (5) due to $n_1 = n_3$) leads to renormalization of the longitudinal relaxation time [26]. E.g., at $W_F \gg W_S$, $W_P$ we find $T_{1S}^{(P)} \approx 1/W_F \ll T_{1S}^{(0)}$. Further we will omit the upper index for $T_{1S}^{(P)}$.

Let us introduce a dynamical variable $\bar{M}$, which is proportional to the average SE intensity in the solid-state resonator of a phaser generator

$$\bar{M} = 2\rho'\Omega_S (\bar{U})^2 \hbar^{-1} \equiv 2\bar{Y}_S / B', \tag{9}$$

where $B' = T_{2S}\Omega_S |F_{12}|^2 / \rho' V_u^2 \hbar$; $\rho'$ is the crystal density, and the line above denotes averaging over the FPAR volume. On the basis of the wave equation for hypersound in active paramagnetic medium [26], using approximations of works [4], [5], we obtain an equation for the first derivative of phaser SE intensity

$$\frac{d\bar{M}}{dt} = B'\bar{M}\,\bar{N} - \frac{\bar{M}}{T_{1S}}; \tag{10}$$

where $\bar{N} = \overline{(\Delta N)}$.

Averaging (8) as well, we obtain the second equation of our system

$$\frac{d\bar{N}}{dt} = -B'\bar{M}\,\bar{N} + \frac{\bar{N}_0 - \bar{N}}{T_{1S}}; \tag{11}$$

The system (10) - (11) is a system of reduced movement equations for an autonomous phaser. This system, due to reduction of the pumping equations, is isomorphic to the movement equations for a two-level autonomous laser of class "B" [19]. Introducing $\tau = t/T_{1S}$, we will further work with a dimensionless form of these equations. For convenience, let us use the following dynamic variables:

$$J(\tau) = B'T_{1S}\bar{M} \equiv 2\bar{Z}_S\,; \quad n(\tau) = \bar{N}/\bar{N}_{tr} \tag{12}$$

and the following CP :

$$A = \bar{N}_0/\bar{N}_{tr}\,; \quad B = T_{1S}/T_{CAV}, \tag{13}$$

where $\bar{N}_{tr}$ is the threshold value of the inverted population difference of spin levels corresponding to self-excitation of the phonon SE:

$$\bar{N}_{tr} = \frac{1}{B'T_{CAV}} = \frac{\rho' V_u^2 \hbar}{T_{CAV} T_{2S} \Omega_S |F_{12}|^2}. \tag{14}$$

Let us now introduce a periodic perturbation of amplitude $k_m$ and frequency $\omega_m$ into the equation system (10), (11). The concrete form of the perturbation naturally depends on the choice of the modulated CP. As it has been shown theoretically [19] and follows from our experiments, the transition from modulation of one parameter to the modulation of another primarily affects the quantitative movement characteristics of the system, while on the qualitative level, behavior of the modulated system remains essentially unchanged. For convenience, we introduce the modulation into the right-hand side of equation (10) (in this case, according to [19], the smallest values of $k_m$ are needed to reach the first critical points). Then, finally, our equations for a non-linear phaser system containing a microwave solid-state resonator with inverted paramagnetic centers acquire the form

$$\frac{dJ}{d\tau} = \Phi_J(J, n) - BJk_m \cos(\tilde{\omega}_m \tau), \tag{15}$$

$$\frac{dn}{d\tau} = \Phi_n(J, n), \tag{16}$$

where $\tau = t/T_{1S}$; $\tilde{\omega}_m = \omega_m T_{1S}$; $B = T_{1S}/T_{CAV}$; $\Phi_J$ and $\Phi_n$ are components of the unperturbed vector field $\vec{\Phi}$ of our system having the following form: $\Phi_J = BJ(n-1)$; $\Phi_n = A - n(J+1)$.

The equation system (15), (16) at $k_m = 0$ has two singular points: saddle [st1]: $(J=0, n=A)$, where $\operatorname{Re}\Lambda_1^{[st1]} > 0$; $\operatorname{Re}\Lambda_2^{[st1]} < 0$; $\operatorname{Im}\Lambda_{1,2}^{[st1]} = 0$ ($\Lambda_{1,2}$ - Lyapunov's indexes) and attractor [st2]: $(J = A-1, n = 1)$, where:

$$\Lambda_{1,2}^{[st2]} = -(A/2) \pm \left[(A/2)^2 - (A-1)B\right]^{1/2}. \tag{17}$$

Taking into attention that $B \gg 1$, at $(4B)^{-1} \ll A - 1 \ll 4B$ we find that [st2] is a focus with so-called relaxation frequency: $\omega_{rel} = \left|\operatorname{Im}\Lambda_{1,2}^{[st2]}\right|/T_{1S} \equiv \tilde{\omega}_{rel}/T_{1S}$, where we introduced, for convenience, the dimensionless frequency $\tilde{\omega}_{rel} = \sqrt{(A-1)B}$.

As it is easy to find for our oscillator (15), (16), at small $k_m$ the branch $B_0$ does softly emerge from this focus as a limiting cycle of period $1T_m$, what has been actually observed in the experiments. The oscillator non-linearity (15), (16) begins to really manifest itself as $k_m$ increases and $T_m$ comes close to the resonance: $T_m \approx T_{rel} \equiv 2\pi/\omega_{rel}$. Here numerical methods are needed. Let us make some necessary estimates by introducing parameters $\varepsilon_1$, $\varepsilon_2$ in the following way:

$$A = 1 + \frac{\varepsilon_2}{\varepsilon_1}; \quad B = \frac{1}{\varepsilon_1 \varepsilon_2}. \tag{18}$$

At $A-1 = O(1)$ we find that relationships $\varepsilon_1 \approx \varepsilon_2 \approx B^{-1/2} \ll 1$; $T_{rel} \equiv 2\pi/\omega_{rel} \approx 2\pi\sqrt{T_{1S}T_{CAV}}$ are fulfilled. At small $A - 1 \approx 4\pi^2/B$ we obtain $\varepsilon_1 \approx 1/2\pi$; $\varepsilon_2 \approx 2\pi/B$; $T_{rel} \approx T_{1S}$. Let us now evaluate the value of $N_{pp}$, which corresponds to the minimum number of required steps with time $T_{step}$ for the modulation period, i.e., $N_{pp} = T_m/T_{step} = 2\pi(\omega_m T_{step})^{-1}$. Accounting for the fact that the field subsystem in our case is fast with respect to the atomic subsystem, for $\omega_m \approx \omega_{rel}$ we obtain

$$N_{pp} > \frac{\kappa_2 T_m}{T_{CAV}} \equiv 2\pi\kappa_2\sqrt{\frac{B}{A-1}} \equiv \frac{2\pi\kappa_2}{\varepsilon_2}, \tag{19}$$

where $\kappa_2 > 1$. E.g., for $A - 1=1/30$ and $B = 3.7 \cdot 10^3$, from (19) we find that $N_{pp} > 2 \cdot 10^3$.

The full evolution time of the system $T_{whole}$ is another important parameter in looking for solutions of our equation system. In all cases, we carried out our trial calculations for such number of modulation periods $N_p \equiv T_{whole}/T_m$ as to surpass the value of $T_{1S}/T_m$ ratio. This means that there should be $N_p \approx \kappa_1/(2\pi\varepsilon_1)$, where $\kappa_1 > 1$. In the vicinity of some special points (e.g., period doubling points) effect of critical deceleration is observed – like at phase transition points in conservative systems. Here $N_p$ values were adapted to the real duration of the transient process by the trial-and-error method. The total number of points for calculations $N_{calc} \equiv N_p \cdot N_{pp}$ outside the critical deceleration regions does not depend on $A$ and has a simple form $N_p \cdot N_{pp} > \kappa_1\kappa_2 B$. However, it should be taken into account that the problem with $A-1 \ll 1$ requires larger $N_{pp}$ because of inverse root dependence of $T_m$ upon the overrun of the generation threshold ($N_p$ is considered to be small while $T_m \approx T_{rel} \to T_{1S}$).

This circumstance is related to the $A$–dependence of the normalized dissipativity $D_n$ introduced in [20] as $D_n \equiv \mathrm{Re}\,\Lambda/|\mathrm{Im}\,\Lambda|$, where Lyapunov's indexes are taken at the stationary point in the absence of modulation. As shown in [20], the value of $D_n$ can be used for preliminary prediction of the degree of phase space stratification – when $|D_n|$ decreases, the number of existing attractors, generally speaking, tends to be increased at the given CP set [20].

For the attracting point [2] of our system in the autonomous mode, using (17), we obtain

$$D_n = -A\left\{2\left[(A-1)B - (A/2)^2\right]^{1/2}\right\}^{-1}, \tag{20}$$

Formula (20) is defined at $A_{NF}^{(-)} < A < A_{NF}^{(+)}$, where saddle-focus bifurcation points $A_{NF}^{(\pm)}$ are $A_{NF}^{(\pm)} = 2B\left\{1 \pm (1-B^{-1})^{1/2}\right\}$. For the phaser system, $B \gg 1$ (typical values are $B \approx 10^3$-$10^5$), so $A_{NF}^{(+)} \approx 4B$; $A_{NF}^{(-)} \approx 1 + (4B)^{-1}$. Then for $A_1 < A < A_2$, where $A_1 - 1 \gg (4B)^{-1}$; $A_2 \ll 4B$, we obtain

$$D_n = \frac{-A}{2\sqrt{(A-1)B}}. \tag{21}$$

As distinct from the unperturbed vector field divergence $\vec{\Phi}_0 \equiv \vec{\Phi}(k_m = 0)$ characterizing the contraction (i.e., the rate of phase volume contraction)

$$\text{div } \vec{\Phi}_0 = \left(n^{[st2]} - 1\right)B - \left(J^{[st2]} + 1\right) = -A \tag{22}$$

the value of $D_n$ has a non-monotonous dependence on $A$:

$$\frac{d}{dA}D_n = \frac{2-A}{2(A-1)^{3/2}}. \tag{23}$$

Instead of linear increasing, which is characteristic for $\left|\text{div } \vec{\Phi}_0\right|$, the normalized dissipativity module is rapidly decreasing when $A$ is changed from 1 to 2, and only then (at $A > 2$) begins to slowly increase with the asymptotics $|D_n| \infty \sqrt{A}$. Consequently, there are three qualitatively different regions of the pumping parameter $A$ (both from physical and calculation point of view) that require separate approaches:

    1. Pumping parameter values are small: $A - 1 \ll 1$, accounting for the aforementioned limitation $A - 1 \gg 1/(4B)$. In this case, slight stratification (layering) of the phase space is predicted (due to large $|D_n|$) in combination with weak contraction (as $\left|\text{div } \vec{\Phi}_0\right|$ is small);

    2. Vicinity of the $|D_n|$ minimum: $A \approx 2$. Here contraction is moderate, but maximum phase space stratification is predicted;

    3. The region of high pumping parameter values, $A \gg 2$ (accounting to be limitation $A \ll 4B$). With increased $A$, a certain decrease in stratification degree is predicted (as compared with the previous case) on the background of enhanced contraction.

Condition 3 is much less typical in real experiments (it is difficult to obtain such high inversion of the spin system) than conditions 1, 2, but we consider all three cases in order to obtain a full picture.

Main results of our numerical modeling of SE dynamics in a solid-state resonator of the phaser generator can be summarized as follows. The total number of the observed coexisting attractors, including limiting cycles with periods $MT_m$ (where $M$ is the external force undertone number), invariant toroids and strange attractors, is first increased with increased $A$ (upon transition from condition 1 to condition 2), and is then decreased upon further monotonous increase of $A$ (transition from condition 2 to condition 3) – according to formula (21). From the other side, the generated SE pulse amplitude is monotonously increasing upon increase of $A$ - in a complete accordance with formula (23). This supports interpretation of $D_n$ and $\text{div } \vec{\Phi}$ as correct characteristics of different behavior aspects of one and the same dissipative system [20].

Let us go now to concrete numerical results. Phase space stratification of our system begins under the scenario of primary saddle-knot bifurcations, which had been already known (e.g., for biochemical systems) in the beginning of 1970-ies [21], and which adequately describes the global properties of Class "B" modulated lasers [19]. Pairs of stable and unstable limiting cycles with periods $MT_m$ ($M = 1, 2, 3, …$), i.e., natural undertones $\omega_m$, evolutionate (with, say, increasing $k_m$) from phase space singularities in all the studied range $A_{low} < A < A_{high}$ ($A_{low} - 1 = 1/30$; $A_{high} = 30$), but the "density" of the coexisting limiting cycles changes non-monotonously in this range.

The simples phase space structure was observed in region 3. The calculations were carried out at $A = 30$; $B = 3700$; $\tilde{\omega}_m = 290$. The primary hard branch $B_1$ is developed into a series of Feigenbaum's period doublings, becoming chaotic. This chaotic attractor at $k_m \approx 0{,}18$ has maximum refractority time $\approx 2T_m$, it coexists with the higher regular branch $B_2$ with initial period $2T_m$, and at $k_m \approx 0.19$ the chaotic attractor belonging to branch $B_1$ is destroyed. As itself, the limiting cycle on branch $B_2$ is not a result of the first step of the doubling series (though its secondary evolution proceeds according to the Feigenbaum's scenario) – this cycle is a primary one, i.e., it is born just with the period $2T_m$ and refractority that is also equal to $2T_m$. The nature of this primary undertone (as all the subsequent primary undertones – see below) is one and the same – continuous formation of independent subharmonic orbits in a dissipative system [21]. This becomes clear with further increase in the modulation amplitude: at $k_m = 0.44$, the only attractor of the system is the limiting cycle with period $3T_m$ and refractority that is also equal to $3T_m$ (for the Feigenbaum's series, the period of $4T_m$ should emerge at once, and refractority should remain equal to $1T_m$). At the same time, for this third hard branch, when $k_m$ is increased from 0.44 to 0.62, a secondary Feigenbaum's evolution is observed, with the tripled initial period: $3T_m \to 6T_m \to 12T_m \to 24T_m \to \ldots$ and the unchanged refractority time $T_R = 3T_m$. For even higher values of $k_m$, chaotic SE mode is realized.

The bifurcation sequences in the region 1 for small $k_m$ are similar to those realized in the region 3. However, upon $k_m$ increase in the region 1 we also observed much more complex dynamic properties. A very small contraction leads to the to excitation of chaotic oscillations with very large refractority time. At $A - 1 = 1/30$; $B = (1/27) \cdot 10^5 \approx 3704$; $k_m = 2.4 \cdot 10^{-2}$; $\tilde{\omega}_m = 10$; $J(0) = 10^{-10}$; $n(0) = 1.0$ we obtain maximum values of $T_R$, up to $5T_m$ (Fig.4). Even higher values - $T_R = 12T_m$ - are noted when the modulation coefficient becomes as high as $k_m = 0.1$ (at the same CP as in Fig.4).

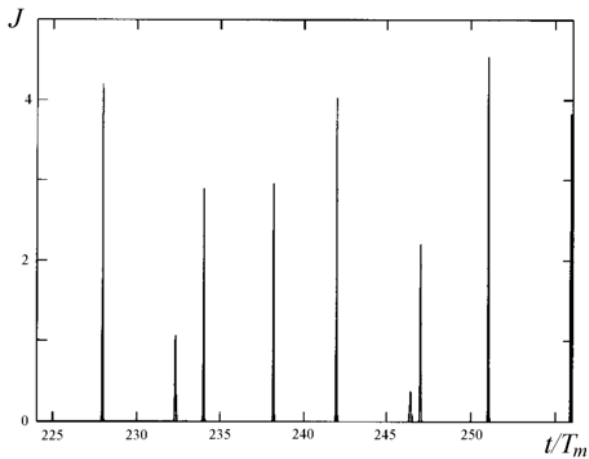 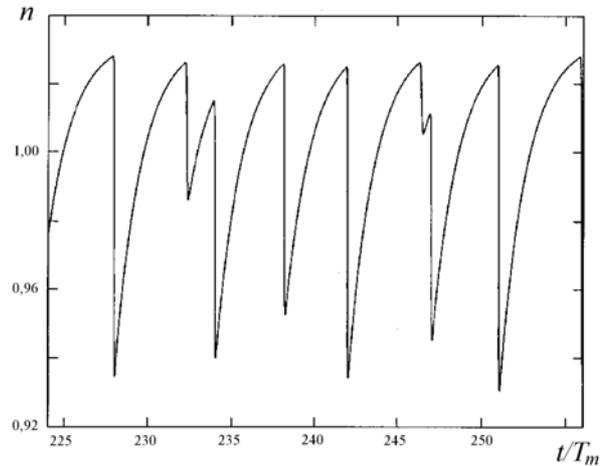

Fig. 4. Calculated time dependence of SE intensity. Refractority, as in the experiment (see Fig. 3, oscillograms b-d), varies within the limits of one to five modulation periods.

Fig. 5. Calculated time dependence of inverse population difference at the same controlling parameters and initial conditions as in Fig. 4.

Comparing the experimentally obtained dependences $J(t)$ for the mode with prolonged variable refractority (oscillograms b-d in Fig.3) with results on $J(t)$ obtained by calculations using our SE model in the region 1 (Fig.4), we can see their good agreement: both in experimental and calculated SE pulse sequences all refractority times are present, without exceptions - from $T_m$ to $5T_m$, including the last value. Moreover, comparing Fig.4 and Fig.5, one can clearly see the qualitative difference in behavior of dynamic variables *J* and *n*, related to substantial difference in relaxation times for the field and the atomic subsystems.

As a matter of fact, the above-described refractority effects are in one or another way related to this substantial difference. As $B \gg 1$ (т. е. $T_{1S} \gg T_{CAV}$), the slow relaxation atomic subsistem (variable *n*) has not enough time "to be tuned" to the fast phonon processes (variable *J*). Therefore, when there is a periodic perturbation imposed from outside, the population difference *n* for certain ranges of CP does not reach over-critical values (ensuring phonon generation) in each of the modulation periods. Correspondingly, it is clearly seen from both experimental (Fig.3, oscillograms 6d) and calculated (Fig.4) pulse sequences $J(t)$ that larger intensity of "irradiated" SE pulse is accompanied by a longer subsequent state of refractority – the phaser is not responding to several modulation pulses. Let us underline once more that in our experiments (where $B \approx 10^4$) such modes are more typical than SE without refractority.

An essential point is that each increase in the maximum refractority time observed in our computer experiments occurred stepwise., was accompanied by an increase in maximum SE pulse amplitude and showed no hysteresis, which was also in full agreement with our real physical experiments.

It is generally assumed [22] that such effects are caused by a qualitative rearrangement of the stratified phase space as a result of so-called external crises (or crises of second kind [22]), when the attraction region of a strange attractor is suddenly expanded, accompanied by an increase in the average refractority time. Each subsequent crisis gives rise to lower and lower pseudoperiods as a result of collisions of the strange attractor with unstable regular manifolds, specifically – with saddle components emerging (at primary saddle-knot bifurcations) together with branches $B_M$ [22]. Therefore, the strange attractor crises can be realized only in these regions of CP space where branching of the above-lying regular attractors has already taken place. We have noted such behavior for all SE modes with pseudoperiods up to $5T_m$ (physical experiment) and $12T_m$ (computer modeling).

As further calculations have shown, when pumping power is increased, and values of $A \approx 2$ are reached, the number of coexisting attractors becomes higher, enhancing stratification of the phase space, which is also in good agreement with experimental data.. The bifurcation diagrams acquire now a very complex indented structure with multiple interweaving in the CP space. As a result, the system sensitivity becomes even higher both to CP perturbations and variation of initial conditions.

As a conclusion, let us stress once more that all the above-described SE processes are adequately described within the deterministic non-linear oscillator model (15), (16) under the double inequality $T_{1S} \gg T_{CAV} \gg T_{2S}$ that is typical for phaser generators. For electromagnetic maser generators, which are also characterized by very low level of intrinsic radiation, condition $T_{CAV} \lesssim T_{2S}$ is normally fulfilled [24], [25]; therefore, for this case a separate consideration is needed, which will be done in our forthcoming papers.

The author is grateful to E.D. Makovetsky, who kindly performed a large part of work on the computer modeling of phaser generation, and to S.D. Makovetskiy, who has written DLL-modules with non-standard library functions for computer experiments.